\documentclass[11pt]{article}
\usepackage{graphicx}
\usepackage{amssymb,amsmath,amsfonts,palatino,amsthm}
\usepackage{amssymb}
\usepackage{epstopdf}
\newcommand{\beq}{\begin{equation}}
\newcommand{\eeq}{\end{equation}}

\newcommand{\f}{\begin{equation}}
\newcommand{\ff}{\end{equation}}

\begin{document}

\title{Classical paradoxes of locality and their possible quantum resolutions in deformed special relativity }
\author{Lee Smolin\thanks{lsmolin@perimeterinstitute.ca} 
\\
\\
Perimeter Institute for Theoretical Physics,\\
31 Caroline Street North, Waterloo, Ontario N2J 2Y5, Canada}
\date{\today}
\maketitle
\vfill

\begin{abstract}

In deformed or doubly special relativity (DSR) the action of the Lorentz group on momentum eigenstates is deformed to preserve a maximal
momentum or minimal length, supposed equal to the Planck length, $l_p = \sqrt{\hbar G}$.  The classical and quantum 
dynamics of a particle propagating in $\kappa$-Minkowski spacetime is discussed in order to examine an apparent paradox of locality
which arises in the classical dynamics.   This is due to the fact that the Lorentz transformations of
spacetime positions of particles depend on their energies, so whether or not a local event, defined by the coincidence of two or more particles, 
takes place appears to depend on the frame of reference of the observer.  Here we discuss two issues which may contribute to the resolution
of these apparent paradoxes.  First it may be that the paradox arises only in the classical picture,  because it is inconsistent to study physics
in which $\hbar =0$ but $l_p = \sqrt{\hbar G} \neq 0$.  Second, there may still be an observer independent notion of a local interaction, which slightly
extends the usual notion without coming into conflict with the observed locality of interactions in nature. 

These considerations may be relevant for phenomenology such as observations by the Fermi observatory, because at leading order in $l_p \times \mbox{distance}$ there
is both a direct and a stochastic dependence of arrival time on energy, due to an additional  spreading of wavepackets.

\end{abstract}
\vfill
\newpage
\tableofcontents

\newpage

\section{Introduction}

Doubly or deformed special relativity (DSR) is the hypothesis that the Poincar\'e group or its action is deformed to take into account the possibility of
a maximal momentum or energy for individual elementary particles, without violating the relativity of inertial 
frames\cite{DSRI,DSRII,DSR-reviews}.  While this is an attractive idea,
not least because it is accessible to investigation by current experiments\cite{GL-phenom}, interpretations of and predictions for these experiments have been
challenged by several confusions as to the interpretation of $DSR$ in spacetime\cite{Unruh-DSR,sabine-paradox}.  These involve the notions of locality and velocity.  The purpose of this paper is to propose
an origin for these confusions which afflict spacetime descriptions of $DSR$ and to investigate three features of $DSR$ theories which may play a role in resolving them.  

One very physical way to understanding the idea of $DSR$ is to see it as a 
phenomenological description which arises in a particular limit of some underlying quantum  theory of gravity\footnote{This viewpoint was first proposed by   Kowalski-Glikman\cite{DSR-reviews}}.  In this limit we take, 
\f
\hbar \rightarrow 0   , \  \  \  \mbox{AND}  \  \  \   G \rightarrow 0
\label{limit1}
\ff
in such a way that the dimensional constant\footnote{In $d=3$, the number of spatial dimensions.}
\f
M_p = \sqrt{\frac{\hbar}{G}} \rightarrow \mbox{constant}
\label{limit2}
\ff
is held fixed.  (We work here in units where $c=1$, since we assume there is still
an invariant velocity.)
That is, we turn off both quantum mechanics and gravity, but in such a way as to preserve phenomena depending on $M_p$ and $c$. 
$DSR$ is then something like the smile that is left of the Chesire cat of quantum gravity.  We can call this the classical $DSR$ limit of quantum
gravity.  Just like the better studied limits in which we take  $\hbar \rightarrow 0$ or $G \rightarrow 0$ separately, this is a limit that must exist if
the quantum theory of gravity is well defined.  

It is easy to state how physics may be modified in the $DSR$ limit: momentum space becomes curved, with the radius of curvature measured by
the invariant $M_p$.  There are then two cases to discuss depending on  how momentum space may be curved.  Poincar\'e  invariance may be broken, in which case the symmetry group of momentum space will have fewer than the ten generators of the Poincar\'e  algebra.  Or, if we want to preserve the existence
of a ten parameter symmetry group, then momentum space must have constant curvature, ie it has a deSitter or anti-deSitter geometry.  This results
in a deformation of Poincar\'e  invariance.  This is the basic idea of $DSR$.  

So long as we stay in momentum space the implementation of this idea is straightforward, but issues develop when we ask for the effect on physics   in the complementary spacetime description.  These problems can be seen to arise because we are working in a limit in which $\hbar$ has been
taken to zero.  In this case there is no  fixed length or time corresponding to the mass fixed in (\ref{limit2}).  Instead,
\f
t_p = \sqrt{\hbar G} \rightarrow 0 
\label{limit3}
\ff
This means that the classical $DSR$ limit only yields something new in 
momentum space.  When applied to physics in spacetime, the classical $DSR$ limit is ordinary special relativity.  

Another way to say this is that the relationship between momentum space and spacetime depends on $\hbar$ being non-zero.  We need it to make
sense of the fourier transform, without it we could not write $e^{\imath \frac{x^a p_a}{\hbar}}$. Alternatively when $\hbar$ is zero $x^a$ and $p_a$ commute, so the idea that one generates translations in the other disappears.  
Hence,  $DSR$ can be only understood as a classical theory in momentum space. If we want to translate the physics of $DSR$ into spacetime we need 
$\hbar \neq 0$ which means we must work with the quantum dynamics.  

In spite of this, there have been attempts to describe $DSR$ physics in the language of classical dynamics of spacetime.  These have 
given rise to some confusion.  Indeed, as pointed out by  SchŸtzhold and  Unruh\cite{Unruh-DSR} and Hossenfelder\cite{sabine-paradox}
there are apparent paradoxes that challenge the consistency of the description.  
They describe thought experiments in which the simple question of whether the world lines of three or more particles coincide at a particular local event in spacetime appears to be observer dependent.  This challenges the basic operational definition of a spacetime event proposed by Einstein, according to which an event is defined by the coincide of several particles  in space and time.  

From the point of view just mentioned, it is not surprising if problems appear when one attempts to discuss the physics of $DSR$ in terms of classical
physics in spacetime, because quantum effects are being treated inconsistently.  If we include
effects coming from a finite $l_p = \sqrt{\hbar G} $ we must include quantum effects because $\hbar$ is non-zero.  This suggests an hypothesis about the 
paradoxes raised in \cite{Unruh-DSR} and \cite{sabine-paradox}, which is that they perhaps arise because it is inconsistent to 
reason about $DSR$ effects in spacetime as if quantum mechanics was turned off, but $l_p$ is still non-zero, because we are being inconsistent about the dependence of observable quantities in $\hbar$.  

\subsection{The strategy of this paper}

It is one thing to suggest an hypothesis about a problem and another to show that it cleanly solves the problem. 
When seeking to go further, however, we run into a problem, which is that the 
scenarios described in \cite{Unruh-DSR} and \cite{sabine-paradox} are discussed in somewhat heuristic contexts.  This is sufficient to convince one there is are issues worth worrying about, but may be insufficient to resolve them, because it is not clear which precise physical theories-if any- correspond to the assumptions that are made there. For better or worse there are several distinct formulations of theories which are motivated by the idea of $DSR$, and it is not known if they are equivalent; nor is it known in all cases if these formulations are completely self-consistent.  There are also two distinct issues that
are easily confused in these discussions:  1)  Is a particular formulation of $DSR$ internally consistent?  2)  If it is consistent, does it lead to predictions that disagree with well confirmed observations such as the locality of physics?

Because of these issues we follow a cautious strategy in addressing the apparent paradoxes raised in \cite{Unruh-DSR} and \cite{sabine-paradox}:

\begin{enumerate}

\item{} We work within the best defined formulation of $DSR$, which is physics on the non-commutative manifold $\kappa$-Minkowksi spacetime.  For simplicity we discuss in this paper only the theory of free relativistic particles on  $\kappa$-Minkowksi spacetime\cite{KP1,KP2,KM1,KM2,previous-free}.

\item{} In this context we formulate a paradox similar to those discussed by  \cite{Unruh-DSR} and \cite{sabine-paradox}.

\item{} We discuss in this paper two approaches to resolving that paradox, one having to do with additional quantum effects special to $DSR$ theories, a second having to do with a possible relaxation of the notion of locality.  

\item{} A third strategy or resolving the paradoxes is discussed in another paper, which is that the apparent non-localities are a coordinate artifact
associated with an ambiguity in extending Einstein's procedure for synchronizing clocks to clocks with a finite frequency far from the origin of
coordinates of a reference frame\cite{me-clockproblem}.

\end{enumerate}

The results are tentative, in that we find evidence that both effects may play a role, but we are not able to definitively show the problem is solved.   It must also be emphasized that because of the strategy we choose we cannot address directly the paradoxes of   Sch\"utzhold and  Unruh\cite{Unruh-DSR} and Hossenfelder\cite{sabine-paradox}\footnote{It should also be mentioned that Hossenfelder\cite{sabine-paradox} does consider the possibility that quantum effects resolve the problem, but comes to a different conclusion then we do here.  Whether that is because the models are different or for other reasons remains unclear.}.  We can here only address similar issues which appear in $\kappa$-Minkowski spacetime; whether the same insights apply to the cases discussed in \cite{Unruh-DSR,sabine-paradox} can only be resolved by further work.  

\subsection{Outline of the argument}
 
This having been said, let us introduce the basic ideas and results that are discussed below.

It is easy to see how an issue with locality arises in  $\kappa$-Minkowski spacetime.  The details are in the sections below but for the sake of clarity of argument we can now sketch the key points of the argument.  

If momentum space is curved then translations on momentum space don't commute with each other.  But if the spacetime coordinates are constructed to be complementary to momentum space in the usual way, they are the generators of translations on momentum space, which means that they don't commute.  So we have a non-commutative
spacetime geometry.  We have then to investigate whether this non-commutative spacetime can support a consistent framework for physics, which agrees with experiment. 

The new commutation relations have to be consistent with a ten parameter algebra of symmetries, inherited from the symmetries of the deSitter geometry of momentum space.  It turns out this can be achieved if we take it that the 
 space coordinates do not commute with the time coordinate 
\f
[t, x^i ] = \imath t_p x^i  
\label{tx0}
\ff
where $t_p= \sqrt{\hbar G}$ is the Planck time.  
This certainly gives rise to some general problems of interpretation because the notion of an event, which requires localizing two or more particles
at the same point of space and time, appears to be compromised by the inability to make both space and time measurements simultaneously sharp\footnote{Alternative formulations of $DSR$ which involve instead of (\ref{tx0}) an
energy dependent metric also give rise to similar confusions, but these will not be discussed here.}.
 
The next step is that, in order to preserve the commutation relations (\ref{tx0}),  the Lorentz transformations of the position and time coordinates of a relativistic  particle turn out to depend on its energy and momentum\cite{DSR-reviews}. (These transformations are reviewed in (\ref{xtrans},\ref{ttrans}) below.) 

This energy and momentum dependence of the Lorentz transformations leads directly to problems with locality. To see why,
consider a scattering  event defined by the coincidence of four worldlines of 
 particles of different energy and momentum-two worldines for the incoming particles and two worldlines for the outgoing particles.  Suppose that one inertial observer sees them as coinciding at a single value of $(x^i,t)$.  Suppose also we want to use a Lorentz transformation to derive the trajectories of those particles as seen by a second intertial observer. Then it is easy to construct
 cases in which the four particles no longer coincide in the second frame of reference, because the modified Lorentz transformation takes the positions coincident in one frame to different locations, depending on the energy and momentum of the particles.    
 
This certainly sounds bad, as it means an interaction that looks local to one observer involves four separated events in another observer's description.  However, we have argued that the classical picture of $DSR$ may not be self-consistent, so this puzzle should be re-examined in the quantum theory. 
To see why quantum effects may help to resolve these apparent paradoxes, note that the 
commutation relation (\ref{tx0}) imply additional uncertainty relations\cite{JKG1,ALN,Rychkov,Tezuka},
\f
\Delta t \Delta x \geq t_p |x|
\label{dtdx}
\ff
This implies immediately that we cannot construct a quantum state in which we can precisely localize a single particle both in  $x^i$ and $t$.  So the first observer cannot actually be sure that the four worldlines coincided at a single event.  All observers must describe the possible interactions amongst the
particles in terms of quantum probability amplitudes.  Indeed, as is discussed by several 
authors\cite{JKG1,ALN,Rychkov,Tezuka,sabine-paradox,GTetal}, this additional uncertainty gives rise to 
 an anomalous spreading of wavepackets
due to the modified commutations relations (\ref{dtdx}).  We reproduce this below, and investigate the extent to which this may provide a solution to the problem of non-locality generated by energy and momentum dependent Lorentz transformations.  

However, we are not able to demonstrate that the spreading of the wavepacket is sufficient to hide the non-locality generated by the Lorentz transformation for all states.  Thus, we next investigate another approach to the issue.  This is whether there might be a relaxation of the notion of locality which can incorporate the Lorentz dependence of the notion of a localized event, without at the same time leading to non-locality being so generic that it blatently disagrees with known physics.  Thus, in section 5 we introduce a notion of $\kappa$-locality in the context of the classical free relativistic particles, which is defined as follows:  {\it A set of $N \geq 3$ events, $E^a_i$, each on a worldline $x^a_i$ of a free relativistic particle, are $\kappa$-local if there is an (energy and momentum dependent) Lorentz transformation which takes the $N$ events to a single event $E^{a \prime}$.}  We can then hypothesize that interactions among particles propagating in $\kappa$-Minkowski spacetime are $\kappa$-local rather than local\footnote{Note that we restrict consideration to $3$ or more particles, because if there is a real interaction amongst two particles, either they exchange energy and momentum, in which case the outgoing worldlines have different energy and momentum than the incoming ones, or they annihilate into a third particle.  Coincidences of two particles that do not change either's energy and momentum are not physical interactions.}.  We can call $N \geq 3$ worldlines that contain mutually $\kappa$-local events, {\it $\kappa$-intersecting.}

This will be acceptable only if generic sets of  $N \geq 3$ worldlines do not contain any mutually $\kappa$-local events.  Equivalently, it should not be the case that $N \geq 3$ sets of worldlines can be brought to intersect by an energy and momentum dependent Lorentz tranformation.  For were this possible, the principle that physics is $\kappa$-local would imply that any set of $N \geq 3$ particles could be interacting.   Were this the case the notion that physics is local would be entirely lost.  

Happily, we show that this catastrophe does not occur.  In section 5 we find that $\kappa$-locality is not a generic property of any set of three or more worldlines.  To the contrary, the sets of $\kappa$-intersecting triples of worldlines are of measure zero in the sets of three worldlines, and become even rarer as four or more particles are involved.  To investigate whether physics was $\kappa$-local rather than local would then take very delicate experiments, with finely tuned initial and final conditions to restrict to the case where the worldlines of the ingoing and outgoing particles are all mutualy $\kappa$-intersecting.  Hence, the world could be $\kappa$-local rather than local,  and we would not yet have noticed the distinction\footnote{We may note that this appears to disagree with the claim of Hossenfelder in \cite{sabine-paradox} that in this language can be translated as the assertion that $\kappa$-local physics is ruled out by experiment.  However, it should be stressed that whether this is due to her model of $DSR$ physics being different or to another cause is not clear at this time.  What we can assert is that no claim can be made that physics in $\kappa$-Minkowski spacetime is grossly in contradiction with the observed locality of physical interactions.}.

There is a third point which may play a role in resolving the paradoxes of locality.  Let us return to the time coordinate $t$.  It is possible to choose
the phase space description so that $t$ commutes with the components of spatial momentum.  Hence $t$ can be used to discuss the 
dynamics in momentum space.  However,  as we shall see in the next section, this comes with a cost, which is that the usual position-momentum commutators are deformed to
 \f
 [x^i , p_j ] = \imath \hbar \delta^i_j f(p_0)
 \ff
 where $f(p_0)$ is the function
 \f
 f(E) = e^{t_p p_0/\hbar}
 \ff
 This has an interesting consequence, which is that we can 
 introduce a new time coordinate that does commute with the $x^i$.  This is\footnote{This is also discussed in \cite{previous-free}.}
\f
T=t + \frac{t_p}{\hbar f(p_0)} x^ip_i
\label{Tdef}
\ff
 It is easy to confirm that,
 \f
 [T, x^i ] =0
 \ff
 Hence, it is unproblematic to discuss events defined in a spacetime coordinatized by $(x^i, T)$, even in the quantum theory.  
 {\it So it is natural to propose that the classical spacetime in which we observers make measurements is defined by clocks that
 measure $T$ and rulers that measure $x^i$.    }
 
 However, $T$ no longer commutes with $p_i$, instead
\f
 [T, p_i ] = \imath t_p p_i
 \label{Tp0}
\ff
Hence there is a good time coordinate, $T$,  to discuss evolution of wavefunctions in position space.  And there is a good time coordinate, $t$, 
to discuss evolution in momentum space.  But they are related to each other by the non-local transform (\ref{Tdef}).  This plays 
a role in the derivation of the spreading of wavepackets in $(x^i, T)$ space.  

While the aim of this paper is to address the issue of physical consistency of $DSR$, we note that there are phenomenological implications
of the results derived below.  By studying the propagation of a wavepacket in $(x^i,T)$ spacetime we are able to study the question of
the energy dependence of the speed of massless particles.  We find that there is a first order variation of the speed of light with energy.
In addition,  there is another first order effect, which is the new contribution to spreading of wavepackets.
This gives a stochastic variation of arrival times of photons proportional to $T t_p \Delta E$, where $T$ is the time traveled and
$\Delta E$ is the uncertainty in energy of the wavepacket.  This is at the same order as the linear dependence of velocity with energy,  and so might be observable in current observations by Fermi and 
other observatories. A new strategy to bound or measure this kind of stochastic effect in the Fermi data needs to be developed. 

The remainder of this paper is organized as follows.  
In the next section we describe a general approach to deforming the quantum physics of a free particle in Minkwoski spacetime and then show
how it can be specialized to $\kappa$-Minkowski spacetime.  In section 3 we show how paradoxes of locality can be generated by studying
inconsistently classical physics in $\kappa$-Minkowski spacetime.  In section 4 we investigate the extent to which these apparent paradoxes  of locality may be resolved in the quantum theory of a free relativistic 
particle\footnote{We note that the uncertainty relations and the resulting spreading of wavepackets have been discussed early in the literature on
$\kappa$ Minkowski spacetime and $DSR$\cite{JKG1,ALN,Rychkov,Tezuka}.  What is new here is only the suggestion  that these may be necessary
to resolve the apparent paradoxes of locality arising from the dependence of boosts of spacetime coordinates on energy. }.   Then in section 5 we introduce the concept of $\kappa$-locality and $\kappa$-intersecting and show that they are very non-generic properties of sets of worldlines.  
We conclude by listing some of the open issues that remain to be resolved before it can be asserted that $DSR$ is either can or cannot be
fully realized within quantum physics and hence before its experimental implications can be unambiguously predicted.

\section{$DSR$ for a free particle in terms of non-commutative geometry}

\subsection{An algebraic approach to DSR}

We begin by examining how the Hilbert space for a single free relativistic particle can be deformed consistently.  We start with a set of 
possible deformed commutation relations. 

\f
[x^i,p_j ]=  \imath \hbar \delta^i_j f(p_0),   \ \ \ \ \ \ \ \  [t, p_0 ]  =  {\imath \hbar} g(p_0) 
\label{CCR2} 
\ff
\f
[t,x^i ] =  \imath t_P x^i h(p_0)   
 \label{tx2}
\ff
with the rest vanishing.  
In particular, we assume,
\f
[t,p_i ] =   0  
 \label{tp2}
\ff
because I would like to define the quantum evolution in a time that commutes with momentum so I can evolve states on 
momentum space in the usual way.

By checking the Jacobi relation
\f
0 = [t,[x^i,p_j ]] + ...
\ff
we find that
\f
\frac{f^\prime}{f}= \frac{t_P}{\hbar}\frac{h}{g}
\ff
It is important to note that the nonvanishing of (\ref{tx2}) follows from the deformations in (\ref{CCR2}) by the Jacobi relations.  So a non-commutativity of
space and time coordinates is a natural consequence of the deformation of the canonical commutation relations for a relativistic particle.  What this
means is that we cannot speak of events or evolve position space wavefunctions in the usual way, so long as we use the time coordinate $t$.  

\subsection{Review of $\kappa$-Poincar\'e}

Let us first briefly review physics in $\kappa$-Minkowski spacetime\cite{KP1,KP2,KM1,KM2,previous-free,DSR-reviews}.
The basic idea is that
momentum space is a deSitter spacetime  coordinatized by $k_0$ and $k_i$ with a metric
\f
ds^2 = -dk_0^2 + e^{\frac{2k_o}{E_p}} dk_i dk_i
\ff
The commutation relations are
\f
[x^i,k_j ]=  \imath \hbar \delta^i_j,   \ \ \ \ \ \ \ \  [t, k_0 ]  =  {\imath \hbar} 
\label{CCR3} 
\ff
\f
[t,x^i ] =  \imath t_p x^i   
 \label{tx3}
\ff
\f
[t,k_i ] =  - \imath t_p  k_i   
 \label{tp3}
\ff
In particular, note that unlike what we have assumed above, the commutator of $t$ with $k_i$ is non-zero.

The dynamics is defined by a Hamiltonian constraint constructed from the Casimir of the $\kappa$-Poincar\'e algebra.  This is
the invariant length on the curved momentum space, which is invariant under an $SO(1,3)$ subgroup of the deSitter group.
\f
{\cal H}= 4E_p^2 sinh^2 \left ( \frac{k_0}{2E_p}  \right )  - k_i k_i e^{\frac{k_0}{E_p}} -m^2 =0
\ff

An integration measure on momentum space, invariant under the non-linear action of the Lorentz group is  defined by 
 \f
d\omega =   dk_0 \wedge  d^3k e^{\frac{3 k_0}{E_p}}  
\label{measurek0ki}
\ff

\subsection{Correspondence with $\kappa$-Poincar\'e}

To construct the quantum theory we prefer to work with our original ansatz according to which $[t,p_i ] =   0  $ (\ref{tp2}).  This way
we can evolve the eigenstates of momentum $p_i$  in the time $t$.   We can see that this corresponds
to the solution of the Jacobi relations given by 
\f
f=e^{\frac{k_0}{E_p}} , \ \ \ \ \  h=g=1
\ff
with the relation
\f
p_i = e^{\frac{k_0}{E_p}} k_i   , \ \ \ \ \   p_0= k_0
\ff
The commutation relations are then
\f
[x^i,p_j ]=  \imath \hbar \delta^i_j e^{\frac{p_0}{E_p}},   \ \ \ \ \ \ \ \  [t, p_0 ]  =  {\imath \hbar} 
\label{CCR4} 
\ff
\f
[t,x^i ] =  \imath t_P x^i   
 \label{tx4}
\ff
\f
[t,p_i ] =   0  
 \label{tp4}
\ff
In terms of these variables the metric on the de Sitter momentum space is
\f
ds^2 = -dp_0^2 (1-\frac{p_i p_i}{E_p^2}  )  -  \frac{p_i dp_i dp_0 }{E_p}     +    dp_i dp_i
\ff

The Casimir of $\kappa$-Poincar\'e   now takes the form,
\f
{\cal H}= 4E_p^2 sinh^2 \left ( \frac{p_0}{2E_p}  \right )  - p_i p_i e^{-\frac{p_0}{E_p}} -m^2 =0
\label{H-Ep}
\ff
and the invariant measure is
 \f
d\omega =  dp_0 \wedge  d^3p  
\label{measureEP}
\ff

Because of (\ref{tx4}) we cannot discuss evolving the position of the particle in the time $t$.  However, because of (\ref{tp4}), the time $t$ is suitable for evolving the particle in momentum space. 

To evolve the system in the position coordinates $x^i$ we define a new time variable (\ref{Tdef}) which obeys 
\f
[T, x^i ] =0
\ff
However, 
\f
[T, p_i ] = \imath t_P p_i 
\label{[T,p]}
\ff
so $T$ cannot be used to evolve wavefunctions in momentum space parameterized by $p_i$.  

Thus, if we take $p_i$ for the spatial momenta, we see the very interesting conclusion that the evolution in position space and momentum space must take place with different time
variables, whose relation to each other, as defined by (\ref{Tdef}), is non-local in both position and momentum space.  

We can however eliminate (\ref{[T,p]}) by going back to measuring spatial momenta in terms of $k_i$, because we have,
\f
[T, k_i ] =0
\label{[T,k]}
\ff

\section{Classical paradoxes of locality and their quantum resolutions}

We now show how paradoxes of locality can be generated by inconsistently taking $\hbar =0$ but $t_p\neq 0$.  Then we show how
they may be resolved when $\hbar$ is turned back on.

\subsection{Non-locality}

We can see just from the algebra of observables that there will be apparent issues with non-locality if we use the wrong set of coordinates.  Suppose that 
we subject our particle to sudden force coming which is local in space so it occurs at a particular $x^i=a^i$.  
Since $x^i$ and $T$ commute we can localize the event precisely also in $T$, so that it occurs at a particular $T=T_a$.
Thus, in 
the $(x^i, T) $ variables, the force can be modeled as coming from from a potential 
\f
V(x,T)= \delta^3 (x^i -a^i )\delta (T-T_a )
\ff
Note that we could not write a potential local in terms of $x^i$ and $t$ because they don't commute.  Hence, 
\f
V^\prime (x,t)= ?  \delta^3 (x^i -a^i )\delta (t-t_a )
\ff
is undefined as it is beset with operator ordering issues.  

So let us stick with the first event, local in $x^i$ and $T$, but suppose we want to describe when it happens in terms of 
 the other time variable, $t$.  We will  have that it takes place at $a^i$ but at a different, momentum dependent time, given by
\f
t=T_a - \frac{t_P}{\hbar f (p_0) } a^i p_i
\label{Tconsequence}
\ff
however, we cannot measure $T$ and $p_i$ at the same time,  since $[T,p_i ]$ doesn't vanish.  Alternatively, since $x^i$ is sharply
defined, $p_i$ is maximally uncertain.  So we cannot predict when the
event will take place in the $t$ coordinate.   

Conversely, if we measure $p_i$ we can determine that the event takes place at a definite $t-t_a$ with a particular
momentum $b_i$.  But then in terms of the spacetime variables, the event will take place at a position dependent time
\f
T=t + \frac{t_P}{\hbar f } x^i b_i
\label{Tconsequence2}
\ff
but since $p_i$ has been measured sharply, $x^i$ will be maximally uncertain, so when the event takes place in terms of $T$ will
be maximally uncertain.  

These facts have to be taken into account carefully in any description of events in spacetime.  To investigate their effect on propagation of particles we will in the next section  construct the quantum theory of a single relativistic particle.  But first we see how the paradoxes
we referred to in the introduction arise.

\subsection{Boosts and events}

The issues that give rise to the apparent paradoxes of Unruh et al and Hossenfelder become apparent when one writes down the Lorentz
transformations for position in $\kappa$-Minkowski spacetime.  From \cite{JM1,JM2} we find for a boost of magnitude $\beta \gamma $ denoted by a
spatial vector $\omega^i$, the position and time coordinates transform as 
\f
\delta x^i = - \omega^i t - t_p  \epsilon^{ijk}\omega_j L_k
\label{xtrans}
\ff
\f
\delta t = -\omega \cdot x + t_p \omega \cdot N
\label{ttrans}
\ff
where $L_i$ are the spatial angular momentum generators
\f
L^i = \frac{1}{f} \epsilon^{ijk} x_j p_k
\ff
and $N_i$ are the generators of deformed boost transformations
\f
N_i =- p_i e^{-\frac{p_0}{E_p}} t - x_i \left [ \frac{E_p}{2} (1- e^{-\frac{2p_0}{E_p}} ) + \frac{1}{2E_p}p\cdot p e^{-\frac{2p_0}{E_p}} \right ]
\label{boosts}
\ff

One can also check that
\f
\delta T = - \omega \cdot x (1  +  t_p r(p_0)) + t_p \omega \cdot p e^{-\frac{p_0}{E_p}} \left (  T-\frac{ t_p}{\hbar} x^i p_i e^{-\frac{p_0}{E_p}}  \right )
\label{Ttrans}
\ff
where 
\f
r(p_0) = \frac{1}{2t_p} (1 - e^{-2 t_p p_0}) + t_p p \cdot p e^{-2 t_p p_0} \approx p_0 + ...
\ff

To first order in $t_p$ we have
\f
\delta T = - \omega \cdot x (1  +  t_p p_0) + t_p \omega \cdot p  T + O(t_p^2) 
\label{Ttrans2}
\ff

We now see how some apparent paradoxes stem from these transformation laws.

\subsection{Transverse length contraction and relativity}

We note first of all, that from  (\ref{xtrans}),  directions perpendicular to the direction of the boost can contract.  This gives rise to an apparent paradox.
Consider two inertial observers, Alice and Bob,  who are approaching each other along their $\hat{z}$ axes, each of whom carries a stick along their $\hat{x}$ axis. 
Suppose that by  (\ref{xtrans}) Alice sees Bob's stick  contract.  What does Bob see?  By the relativity of inertial frames Bob should also see Alice's
stick contract.  This is what we say with sticks parallel to their relative velocities and this turns out to be consistent.  But now let us note that as they pass
Alice and Bob can mark where the end of the other's stick passes their stick.  This does not yield a problem when they are held parallel because of the relativity of simultaneity but it is a problem now, because the events of marking the ends of the sticks are simultaneous in both observer's frames.  Hence,
if Alice sees Bob's stick to have contracted relative to hers, Bob must agree that Alice's stick is longer.  But this appears to violate the relativity of inertial
frames.  This is why ordinarily we do not have contraction of directions perpendicular to motion in special relativity.  

The resolution of this problem is that the contraction of the perpendicular directions is proportional to conserved quantities, which in this case is
the $\hat{y}$ component of angular momentum of the sticks.  So whether it is Alice of Bob who sees the other's stick as shorter than theirs depends
on which stick has a larger $\hat{y}$ component of angular momentum.  

\subsection{A classical locality paradox}

There are other more serious apparent paradoxes connected with the fact that the transformations (\ref{xtrans}) and (\ref{ttrans}) are dependent
on energy, momentum and angular momentum of the objects which are being transformed.  

Here is a prototype of an apparent paradox involving transformations between the observations of two inertial observers Alice and Bob.  

Suppose that Alice sees a collision of two particles
at a position $a^i$ and time $T= s$ in her frame.  We can use the  transformations (\ref{xtrans}) and (\ref{ttrans}) to compute the first order
positions and times of the particles at the collision, as they will be seen according to Bob's instruments.  Bob will see the two particles to have
positions and times given by
\f
x^{i \prime} = a^i + \delta x^i,    \ \ \ \ \   T^\prime = s +\delta T
\ff
where $\delta x^i$ and $\delta T$ are given by (\ref{xtrans}) and (\ref{Ttrans}), respectively.  

If the two particles have different values of energy, momentum and angular momentum, Bob will see the event that Alice sees as two particles colliding as corresponding to two events, separated
in space by a vector with space and time components $D x^i$ and $DT$,  given by 
\f
D x^i =  t_p \left (  \omega^i  a^k D p_k  -  \epsilon^{ijk}\omega_j D L_k  \right )
\label{differencex}
\ff
\f
D T =- t_p \left (    \omega \cdot a D E  -s \omega \cdot \Delta p    \right )
\label{differencet}
\ff
where $D p_k, DE $ and $D L^i$ are the differences in conserved quantities carried by the two particles, as
observed by Alice, ie $Dp^k = p_1^k-p_2^k$, etc. 

Thus, the two particles that Alice sees coincide need not be seen to coincide by Bob. Indeed, since there is a shift perpendicular
to the direction of the boost it is possible that they never collide at all.  It is thus easy to construct apparent paradoxes by, for example,
supposing that the two particles have a short ranged interaction that scatters them.  Suppose that the boost to Bob's frame is
large enough that the $D x$ are larger than the interaction range.  Does Alice see them to scatter, but Bob not?

\section{Quantum theory of a free relativistic particle}

The purpose of this section is to investigate whether quantum dynamics can contribute to the resolution of the puzzle we have just discussed.  

We study the quantum dynamics of a free particle in $\kappa$-Minkowski spacetime. and show that there is an anomalous
spreading of the wavepacket proportional to $t_p \times \mbox{distance} $.  This shows that quantum effects could contribute to the 
resolution of the locality paradox.

\subsection{Bases in Hilbert space}

We begin our study of quantum physics in $\kappa$-Minkowski spacetime by constructing the Hilbert space, $\cal H$, by constructing eigenstates of complete commuting sets of operators.  

We will see shortly that a key point of our approach is that the dynamics is defined first in momentum space, then transformed to the commutative spacetime
defined by the $(x^i,T)$ operators. 

One complete commuting set of observables is composed of $(E, p_i)$.  They define a basis
\f
\hat{p}_i |E, p> = p_i |E, p>, \ \ \ \ \ \ \ \hat{E} |E, p> = E |E, p>
\ff
with completeness relation given by the measure (\ref{measureEP})
\f
1= \int dE d^3p |E, p><E, p|
\label{completeEP}
\ff
 
Another complete commuting set is $(t, p_i)$.   These are useful for discussing evolution in time in momentum space.  These define a basis 
\f
\hat{p}_i |t, p> = p_i |t, p>, \ \ \ \ \ \ \ \hat{t} |t, p> = t |t, p>
\ff
with completeness relation
\f
1= \int dt d^3p |t, p><t, p |
\label{completetp}
\ff
We note that because of $[t, E ]  =  {\imath \hbar} $ we have
\f
<t, p |E , p^\prime > = \delta^3 (p,p^\prime ) e^{\imath Et/\hbar}
\label{Et transform}
\ff

We note that there is no basis which simultaneously diagonalizes $t$ and $x^i$.  

If we want to discuss evolution in position space we have to use a different  time coordinate, $T$, which is part of another complete set of commuting observables, given by $(T, x^i)$.  They define a basis by
\f
\hat{x}^i |T,x> = x^i |T,x>, \ \ \ \ \ \ \ \hat{T} |T,x> = T |T,x>
\ff
with completeness relation
\f
1= \int dT d^3x |T,x><T,x |
\label{completeTx}
\ff

To transform from an energy basis to evolution in position space then requires two steps.  First we have to change from energy to the
time $t$ by using the relation (\ref{Et transform}).  

Next we change from the $(t,p)$ basis to the $(T,x)$ basis through amplitudes
\f
<t,p|T,x> = e^{\imath\frac{ x^i p_i }{f(E) \hbar}  } \delta ( T-t -\frac{t_P}{\hbar f } x^i p_i    )
\ff

This gives us the non-local transform
\begin{eqnarray}
\Psi (x^i,T) &=& < T, x | \Psi >  
\nonumber  \\ 
&=& \int d^3p dt d^3p^\prime dE   < T, x |  t,p><t,p | E, p^\prime >  < E, p^\prime   |\Psi >   \nonumber \\
&=& \int d^3p  dE  
e^{-\imath\frac{ x^i p_i}{\hbar f(E)}  + \imath E  (T- \frac{t_p}{\hbar f }x^ip_i )}  \Psi (E,p)
\label{transform}
\end{eqnarray}

\subsection{Dynamics in momentum space}

We will impose dynamics by defining a modified dispersion relation in energy-momentum space.  We define the dynamical 
Hamiltonian constraint operator by the quantization of (\ref{H-Ep}). 
\f
\hat{\cal H}= 4E_p^2 sinh^2 \left ( \frac{E}{2E_p}  \right )  - p_i p_i e^{-\frac{E}{E_p}} -m^2 =0
\label{Cquantum}
\ff
We define the physical Hilbert space ${\cal H}_{Phys}$ to be the subspace of $\cal H$ defined by
\f
\hat{\cal H} |\Psi > =0
\ff
In energy-momentum space the solutions of this are given by
\f
\Psi (E,p) = \delta (4E_p^2 sinh^2 \left ( \frac{E}{2E_p}  \right )  - p_i p_i e^{-\frac{E}{E_p}}  -m^2    )\chi (p) . 
\label{solution}
\ff

\subsection{Dynamics in position space}

\subsubsection{Propagation and spreading of a wavepacket}

We now perform the transform (\ref{transform}) in the case of one spatial dimension with a solution (\ref{solution}) with
\f
\chi (p) = e^{-\frac{(p-p_0 )^2 }{2 \Delta p ^2}}
\ff
With $\Delta x = \frac{f(E) }{\Delta p} \approx \frac{1 }{\Delta p}$ the result is proportional to
\f
\Psi (x,T)  \approx exp \left (- \frac{1}{2 \Delta w^2}  [  (x-T)^2  -8 x l_pp_0 (x-T)   ]    
+\frac{\imath}{1+\frac{4 x^2 l_p^2 }{\Delta x^4}} [ p_0 (x-T) + \frac{2xl_p (x-T)^2 }{\Delta x^4}  ]   \right ) 
\ff
where there is a new width
\f
\Delta w = \sqrt {\Delta x^2 +  x^2 l_p^2 \Delta p^2}
\ff

We see that there is a new effect proportional to $l_p$ in which  the width spreads out as the particle propagates.  

When the particle reaches an $x > \Delta x  \frac{\Delta x }{l_p}$ the width grows as 
\f
\Delta w \approx x l_p \Delta p
\label{Deltaw}
\ff
This is implies there is a stochastic effect on the time of propagation which is linear in $l_p$ and of the same order as the linear variation in velocity.

It is interesting to ask whether this additional wave packet spreading can address the paradoxes of locality we discussed in section 3.4.
We can be sure that the resulting quantum uncertainty would dominate over the locality paradox if in every frame of reference
\f
\Delta w > Dx,    \ \ \ \ \ \   \Delta w > DT
\ff
Neglecting the transverse term proportional to the angular momentum we see that this would be the case if for all particles,
\f
\Delta p > |p|. 
\label{condition}
\ff
This implies, using (\ref{differencex}) and (\ref{differencet}) that
\f
\Delta w \approx x l_p \Delta p > x l_p p  > |\omega | x l_p D p = |Dx|
\ff
Thefore, for states where (\ref{condition}) is true in every frame of reference the quantum uncertanty will dominate over the apparent non-localities
created by the momentum dependence of lorentz transformations.  This is encouraging, but not definitive, for one thing because
there are states for which (\ref{condition}) is not satisfied.  

\subsubsection{Wave equation in spacetime}

We have constructed the wavefunction by transforming from a wavepacket in momentum space, but this should be equivalent to solving
a wave equation in spacetime.  The problem is that the corresponding wave equation, while linear in $\Psi (x,T) $ is a complicated function
of space and time derivatives.   To see what it is, we 
transform the constraint (\ref{Cquantum}) to a wave-equation on spacetime.  Using the transform (\ref{transform})
we find that 
\f
\hat{E} \rightarrow  -\imath \hbar \frac{\partial}{\partial T} ,  \ \ \ \ \ \ \  
\hat{p}_i \rightarrow - \imath \hbar  e^{-\imath t_p \frac{\partial}{\partial T} }\frac{\partial}{\partial x^i}
\ff

So the wave equation gotten by transforming  (\ref{Cquantum})  is
\f
\hat{H} \Psi (x^i, T) =\left (  \frac{4}{t_p^2} \sinh^2 (-\imath \hbar \frac{\partial}{\partial T}) 
-e^{ \imath t_p \frac{\partial}{\partial T} }  \nabla^2 - m^2  \right )  \Psi (x^i, T)  =0
\label{xTequation}
\ff

We see that the spacetime wave equation (\ref{xTequation}) is of infinite order in time derivatives.  This means that while every transform
of a solution to  (\ref{Cquantum}) on momentum space may solve (\ref{xTequation}), the latter may have many more solutions that do not arise
from the transform of a solution on momentum space\footnote{For a contrary view please see \cite{sabine-contrary}.}.  This is another reason to take the momentum space description as fundamental.
We can take the point of view that only solutions to (\ref{xTequation})  that arise from a transform from momentum space
be taken as physically acceptable solutions.  This is a function on a three manifold's worth, which is correct.

\subsection{Velocity in the commutative space-time coordinates $(x^i,T)$}

There has been a lot of confusion as to the definition of velocity of particles in $\kappa-$Minkowski spacetime.  We can see why,   by working out the 
phase velocity as well as the classical point particle velocity from the canonical theory.  We will see that they are not the same.  

\subsubsection{Phase velocity}

It is straightforward to compute the phase velocity from the wavefunction for the wave packet or, equivalently, from the plane wave,
\f
 \Psi_p (x^i, T) = e^{ -\imath [ \frac{x^i p_i e^{-t_p E_p }}{\hbar} -E_p (T-t_p  \frac{x^i p_i e^{-t_p E_p }}{\hbar}) ]      }
\ff
where $E_p$ is given by the solution of (\ref{H-Ep}).

The principle of stationary phase gives, for propagation in the $z$ direction
\f
v = \frac{dz}{dT} = \frac{E_p}{p_ze^{-t_p E_p }(1+ t_p E_p)} = \frac{e^{\frac{t_p E_p}{2}} }{ (1+ t_p E_p )} \approx 1 - \frac{t_p E}{2} + ...
\ff
so we see there is a leading order subliminal dependence of the speed of light with energy.  

\subsubsection{Classical computation of velocity}

It is also possible to define a classical notion of velocity using the evolution generated by the Hamiltonian constraint, $\cal H$.
This follows the standard computation,  (\cite{velocity-first}) for $\frac{dx^i }{dt}$.  However we argue that because $x^i$ and $t$ don't commute, it is more
relevant for the classical limit of the quantum theory to compute the velocity in the commutative spacetime coordinates, $\frac{dx^i }{dT}$.
This is defined by the ratio
\f
\frac{dx^i }{dT} = \frac{dx^i }{d\tau} ( \frac{dT }{d\tau})^{-1}
\ff
where $\tau$ is the arbitrary time parameter which is used to parameterize evolution generated by the Hamiltonian constraint.
\f
\frac{dx^i }{d\tau} = \{ x^i , {\cal N}{\cal H} \} = -2 {\cal N} p_i
\ff
where $\cal N$ is an arbitrary lapse function.  We find also that
\f
\frac{dT }{d\tau} = \{ T, {\cal N}{\cal H} \} = {\cal N} \left ( \frac{\partial {\cal H}}{\partial E} + t_p p_i \frac{\partial {\cal H}}{\partial p_i} 
\right )
=
{\cal N} \left (4 \sinh (\frac{t_pE}{2} )\sinh (\frac{t_pE}{2} ) - t_p^2 p^2 e^{-t_pE }
\right )
\ff
The result after using the Hamiltonian constraint is
\f
|\frac{dx^i }{dT} | = e^{t_p E}
\ff
We note that this is different from the phase velocity computed above.  

It also differs from the result \cite{velocity-first}
\f
|\frac{dx^i }{dt} | =1 
\ff
so whether there is an energy dependent speed of light at the classical level does indeed depend on the definition of time used. 

\section{Redefining locality}


It is clear to begin with that the usual principle of locality cannot be realized in a quantum theory which includes the commutation relations
(\ref{tx0}).  What then must be investigated is whether the notion of locality can be relaxed sufficiently to incorporate (\ref{tx0}) while remaining sufficiently restrictive that it is not ruled out by our existing store of knowledge.   In this section we make such a proposal and argue that it is 
narrow enough to not yet have been ruled out by experiment.

In the following we  we restrict ourselves to the context in which the paradoxes of locality originally presents the paradox, which is the classical dynamics of worldlines of free relativistic particles.  We will consider cases where three or more world-lines intersect at an event, because unless the outgoing states are different from the incoming states we cannot say that an interaction has taken place.  
We thus  consider a class of events involving $N \geq 3$  particles in which one inertial observer, named Alice, sees all $N$ worldlines to intersect at a single point, $E^a$.  For example, two of them may be incoming states and two of them outgoing, in which case $N=4$.  Alice will then explain
the fact that the particles scattered, exchanging energy and momentum, as caused by the two incoming particles's worldlines having intersected.  

However, if $DSR$ is in fact observer independent, then another intertial  observer, Bob, must observe that the scattering take place as well. However, because of the energy and momentum dependence of the Lorentz transfomrations, Bob may see four worldlines that never intersect.  Instead, the event $E^a$ has been split into four events.  Bob will be forced to describe the history Alice sees as one in which there are two incoming particles  which exchanged energy and momentum and became two outgoing particles without any of the four worldlines intersecting.  Thus it would appear that
observer independence requires that we believe that physical interactions are non-local.  

But if any two non-insecting worldlines can exchange energy and momentum, generating two outgoing worldlines, which intersect neither with each other, nor with the incoming worldlines, then it would seem that any two particles in nature can interact with the same probalities of those whose worldlines intersect.  So physics becomes completely non-local.  This is the catastrophe claimed in \cite{sabine-paradox}.  

But we should be careful before jumping to this conclusion.  What is true is that the concept of locality has to be altered to allow for the observer
dependence of the notion that worldlines coincide in spacetime.  But this need not imply that the notion of locality becomes so weakened that
they theory can be ruled out by our store of observations that we usually take to support the postulate that physical interactions are local.
Might it be possible that the concept of locality is weakened or altered in a way that incorporates the observer dependence of worldline coincidence,
but in a way that still restricts physical interactions in a way that is consistent with observations?

To investigate this we have to formulate exactly what is the concept of locality that is allowed by the structure of physics in $\kappa$-Minkoski spacetime.  
In this regard, I propose for consideration the following notion of locality, which I will call $\kappa$-locality.  I define it as follows.  $N \geq 3$  events $E^a_I$, $I=1,2,3...$ each on the worldline of a particle, $x^a_I (\tau)$ are $\kappa$-local if there is a passive Lorentz transformation to the description seen by another observer, which maps the these events to a single event, $p^a$.     ($x^a_I (\tau)= (t,x^i_I (\tau) )$.  Since each event is on a different worldline the Lorentz transformation leads them to intersect at a common point.    Thus we say that the worldlines which do not intersect, but which can be mapped to intersecting worldlines by an energy and momentum dependent Lorentz transformation are $\kappa$-intersecting.  

We then make a postulate which is a consequence of observer independence:  {\it Suppose that there are $N \geq 3$ events, on $N$ worldlines, which are $kappa$-local. Then an interaction will take place with the same probability as it is seen to take place in the frame of the observer who sees the
three events mapped to a single event, which is a coincidence of all three worldlines.  }  We can call this the hypothesis of $\kappa$-locality of physics.  This can be easily generalized to more than three events. Note that, for the interaction probability to be non-zero, in addition to the worldlines coinciding the onservation of energy and momentum must be satisfied.  So the condition of the worldlines being $\kappa$-intersecting is a necessary but not a  sufficient condition for an interaction to take place.  

We next ask whether the hypothesis of $\kappa$-locality could be consistent with observation.
There will certainly be an inconsistency with observation if $\kappa$-locality is generic.  By this I mean the following.  Consider an arbitrarily chosen
$N$-tuplet of  world-lines of particles in spacetime.  Each is a solution of the equations of motion and so is specified by a point in the phase space of a free relativistic particle.  Generically they do not intersect, (we assume for the moment that space has $d=3$ dimensions, below we consider the special case of $d=1$.)
Now we say that $\kappa$-locality is generic if for every such triple of particle worldlines there is an inertial observer who sees them to intersect at a point.
That is, there are $N \geq 3$ events on the $N$ worldlines that are taken to a single event under some momentum dependent Lorentz transformation.  

If $\kappa$-locality is generic in this sense then the postulate of $\kappa$-locality of physics still does not imply that any $N \geq 3$ particles interact just as if they had an intersection.  The conservation laws still have to be satisfied.  But physics would still look very non-local because an interaction would be no more or less likely to happen then for the case of particles whose worldlines intersect at an event.  

But suppose instead that $\kappa$-locality was a property of measure zero among triples of worldlines of particles.  Then the situation is very different.   
In this case, we cannot use the fact that physical interactions have so far been found to be local to rule out the hypothesis that physics is instead $\kappa$-local.  This is because to see that physics is $\kappa$-local rather than local we would have to very carefully prepare triples of particles on worldlines that have the property that a $\kappa$-local set can be made by selecting one event on each worldline.    Since such experiments have never been made we cannot rule out the hypothesis that physics is $\kappa-$local.  Indeed, it might be of interest to investigate how such experiments might be done. 


Below I study separately the cases of $d=1$ and $d=3$ dimensions and show that in both cases $\kappa$-locality is a property of measure zero among triples or higher multiplets of worldlines.  Hence I conclude that at least in $\kappa$-Minkowski spacetime the argument for a bound from observation fails.  At the end of this section,  I consider the possibility that even if very rare, $\kappa$ local interactions might contradict our understanding of very massive bodies like stars.  The conclusion is that we cannot argue from existing experiments that we not live in $\kappa$-Minkwoski spacetime.  

\subsection{Demonstration of non-genericity of $\kappa$-intersection for $3$ or more  particles in  $d=3$ spatial dimensions}

I work here with the non-commuting coordinates $(x^i,t)$\footnote{It is easy to verify that the same conclusions apply to the case of intersecting worldlines in the commuting spacetime $(x^i, T)$.}.   In one observer's frame, called Alice, three particles described by classical worldlines coincide, an incoming photon, an incoming electron and an outgoing electron.  From this local event we can certainly construct $\kappa$ local triples of events.  Under a Lorentz transformation to Bob's frame the event $E$ is taken to three separated events, $E\prime_\alpha$, $\alpha=1,2,3$ which are the image of $E$ under the energy dependent Lorentz boosts for each of the particles.   
 
So we can create $\kappa$-intersecting triples of worldlines by Lorentz transformations in this way from intersecting triples of worldlines.  The question is, are such $\kappa$ intersecting triples of worldlines,  generic?  That is given any three worldlines, $x^a_\alpha (\tau_\alpha) $   in $3+1$ dimensional Minkowski spacetime, can we choose a triple of events, one on each of the world lines, $E^a_\alpha =  x^a_\alpha (\tau_\alpha)$ so that there is  an energy dependent Lorentz transformation given by (\ref{xtrans}) and (\ref{ttrans}), below, that brings them to one event.

\subsubsection{The set up}

We can choose the gauge in which for each particle, 
\f
x^0_\alpha (\tau) = t_\alpha (\tau ) =\tau_\alpha
\ff
Then the trajectories are given in terms of  the initial conditions $(z^i_\alpha , p_{i \alpha })$ by
by
\f
x^i_\alpha (t_\alpha ) = z^i_\alpha + t_\alpha p_{i \alpha }, 
\ff
After a Lorentz transformation to Bob's frame the trajectories are given by
\f
x^{i \prime} _\alpha (\tau _\alpha ) = x^{i} _\alpha (\tau _\alpha ) + \delta x^i_\alpha  =   z^i_\alpha + t_\alpha p_{i \alpha } + \delta x^i_\alpha  
\ff
Similarly, the new time components are
\f
t_\alpha^\prime  (\tau_\alpha )= t_\alpha (\tau_\alpha  ) + \delta t_\alpha (\tau_\alpha )
\ff

Here $ \delta x^i_\alpha $ and $ \delta t_\alpha $are  functions of the three boost parameters $\omega^i$ plus the initial conditions $(z^i_\alpha , p_{i \alpha })$, by
(\ref{xtrans}) and (\ref{ttrans})

These expressions (\ref{xtrans},\ref{ttrans}) refer to Lorentz boosting around the origin of coordinates.  But we can first do a translation,
then boost.   In $\kappa$-Poincare there is also a deformation of the action of translations, these are given by \cite{JM1,JM2},
\f
\delta t= a^0 + t_p p_i a^i e^{-t_p p_0 } ; \ \ \ \ \ \ \ \delta x^i = a^i
\label{translate}
\ff
for infinitesimal translations labeled by a four vector, $(a^0, a^i)$.  

If we combine a translation of the origin with a Lorentz boost we have $\delta x^a = (\delta t, \delta x^i )$, given by
\f
\delta x^i = - \omega^i (t + a^0 + t_p p \cdot a e^{-t_p p_0 } )  - t_p  \epsilon^{ijk}\omega_j ( L_k - \epsilon_{klm}a^l p^m )
\label{xtrans2}
\ff
and 
\f
\delta t =  -\omega \cdot (\vec{x} + \vec{a}) [1 + t_p E]+  t_p p \cdot a e^{-t_p p_0 } (t + a^0 + t_p p \cdot a e^{-t_p p_0 })    O(t_p^2)
\label{ttrans2}
\ff
We note that rotations add nothing as they do not depend on energy or momentum.  

\subsubsection{Bringing three or more worldlines to intersect}

We start with the case of three generic worldlines and ask whether there exists generically a Lorentz transformation that will bring them to
intersect. Once we have seen that this can't generically be done we will see that as we increase the number of worldlines the constraints 
also increase.

We want to know if generically there exist three points, one on each wordline, $E^a_\alpha = x^a_\alpha (t_\alpha)$ such that there is a Lorentz transformation that brings them together to a single event.  Lorentz transformations are parameterized generically by a boost parameter $\omega^i$
and a translation $a^a$.  Given these there are three difference vectors
\f
\Delta x^a_{\alpha \beta} = E^a_\alpha - E^a_\beta
\ff
of which any two, say $\Delta x^a_{12} $ and $\Delta x^a_{13} $, are independent. 
Let us define the results of Lorentz transformations to be
\f
\delta x_{\alpha \beta}^a= \delta x_{\alpha}^a- \delta x_{\beta}^a, 
\ff
If the three points are brought together then 
\f
E^a_\alpha + \delta x_{\alpha}^a =  E^a_\beta + \delta x_{\beta}^a
\ff
or 
\f
\Delta x^a_{\alpha \beta} = \delta x_{\alpha \beta}^a
\label{Ddelta}
\ff

\subsubsection{A simplifying assumption}

First we run the argument with a simplifying assumption.  Below we will see the result is the same if we drop it, so this is just to make the reasons
for the result clear.

Let us suppose that there is a Lorentz frame in which the three $E^a_\alpha$ are simultaneous.  Let us first transform to that frame of reference.
This means that the three $t_\alpha$ are equal to each other and hence all equal to a single time, $s$.  Hence, the three $\Delta t_{\alpha \beta}=0$.

It follows that
\f
\delta x_{\alpha \beta}^i=  - t_p  \epsilon^{ijk}\omega_j ( \Delta L_{k \alpha \beta}  - \epsilon_{klm}a^l \Delta p^m_{\alpha \beta} )
-t_p \omega^i a_k \Delta p^k_{\alpha \beta}
\ff
Here $ \Delta L_{k \alpha \beta} =  L_{k \alpha }- L_{k \beta }$ and similarly for $\Delta p^i_{\alpha \beta}$.  

We also have
\f
\delta t_{\alpha \beta} =  -\omega \cdot (\vec{x}_\alpha [1 + t_p E_\alpha ]- \vec{x}_\beta [1 + t_p E_\beta ]  + \vec{a}  t_p [E_\alpha -E_\beta ])
+ t_p \omega_i (p^i_\alpha - p^i_\beta     )s
\label{ttrans3}
\ff
Now we notice that 
\f
\omega_i \delta x_{\alpha \beta}^i = -t_p \omega \cdot \omega a_k \Delta p^k_{\alpha \beta}
\label{omegadelta}
\ff
Let us at first consider that the velocity of the boost is also small so we can neglect $\omega \cdot \omega$ and work only to
leading order in $t_p |\omega |$.  Then to this order, we can take 
from (\ref{Ddelta})
\f
\omega_i \Delta x_{\alpha \beta}^i = O(t_p|\omega |^2 ) 
\ff
This means that to leading order in $t_p |\omega |$, $\omega^i$ is normal to the plane formed by $\Delta x_{12}^i $ and $\Delta x_{13}^i $.  Thus, 
to this order, $\omega^i$ can have
just a single component.  If the unit vector proportional to that plane is $\hat{n}^i$ then $\omega^i = \theta \hat{n}^i$ for some $\theta$.

Let us now count the parameters to be varied and the equations to be solved.  We are given the three worldlines, which fixes the trajectories
(in the frame where all the $E^0_\alpha$'s are equal) and the momenta and energies.  For each time $s$  these fix the $ \Delta x_{\alpha \beta}^i $ We had
originally to solve eight equations, six of the form
\f
\delta x^i_{12} = \Delta x_{12}^i ,\ \ \delta x^i_{13} = \Delta x_{13}^i 
\label{tosolvex}
\ff
and two more
\f
\delta t_{12}=0, \ \ \ \  \delta t_{13}=0,
\ff
However, our simplifying assumption has by (\ref{omegadelta}) already solved one component each of the two equations in (\ref{tosolvex}). Hence,
to the order we are working, there
are six more equations to solve.  

The $\delta^i_{\alpha \beta}$ and $\delta t_{\alpha \beta}$ are functions of one $\theta$ and three translation components, $a^i$.  Since we are
varying also $s$ that gives us $5$ degrees of freedom, to vary to look for a simultaneous solution to $6$ equations.  This is one more equations than
variables to vary so if there are any solutions they are highly non-generic. 

\subsubsection{Dropping the simplifying assumption}

Now we show we can drop the simplifying assumption and get the same answer.  So we do not assume that the events are to begin
with simultaneous and we also include terms of order $t_p |\omega |^2 $.  

We now have to solve in place of (\ref{tosolvex}), a more complicated expression
\f
\delta x_{\alpha \beta}^i=  - \omega^i (t_\alpha - t_\beta + t_p a_k \Delta p^k_{\alpha \beta})   - t_p  \epsilon^{ijk}\omega_j ( \Delta L_{k \alpha \beta}  - \epsilon_{klm}a^l \Delta p^m_{\alpha \beta} ) = \Delta x_{\alpha \beta}^i
\label{tosolvex2}
\ff
Then there will be components of $\omega^i$ besides $\theta \hat{n}^i$.  Let us then write 
\f
\omega^i = \phi_{12} \Delta x_{12}^i +  \phi_{13} \Delta x_{13}^i+ \theta \hat{n}^i
\ff
From these we can compute the components $\Delta x_{12}^i \omega_i$ and$\Delta x_{13}^i \omega_i$ .  These are functions of the $\phi_{12},\phi_{13}$ and the norms
and dot products of the $\Delta x_{12}^i $ and $\Delta x_{13}^i$, which are already fixed.  But they also have to satisfy
\f
\omega_i \Delta x^i_{12}= -\omega^i\omega_i (t_1 -t_2+t_p a_k (p_1^k - p_2^k )) , \ \ \ \  
\omega_i \Delta x^i_{13}= -\omega^i\omega_i (t_1 -t_3+t_p a_k (p_1^k - p_3^k ))
\label{tosolvexyes}
\ff
These two equations determine the two remaining components $ \phi_{12}$ and $ \phi_{13}$ in terms of the time differences $(t_1 -t_2)$ and $(t_1 -t_3)$.  Once this is done the counting is as follows.  We now have six equations to satisfy, these are the $2$ components of the $2$ equations (\ref{tosolvex}) in the plane orthogonal to $\omega$, 
plus two more equations
\f
\delta t_{12}=t_1-t_2 , \ \ \ \  \delta t_{13}= t_{1}-t_{3} ,
\label{tosolvet2}
\ff
We have five remaining degrees of freedom: $\theta, a^i$ and the one free time, $t_1$ to vary to solve these $6$ equations.  Five free variables is not enough to solve six equations, so again there remain no generic solutions. 

It might help to make a remark about the counting.  Initially we have $8$ equations to solve (\ref{tosolvex},\ref{tosolvet2}).  It seems at first that we have
$9$ free variables with which to solve them.  These are three boost parameters, $\omega^i$, three times $t_\alpha$ and three components of translations
$a^i$ (the time components of translations doesn't come into any of the equations to be solved.)  However in equation (\ref{tosolvexyes}) we see that two of the boost parameters are used up to compensate for the three times being unequal because $\phi_{12}$ and $\phi_{13}$ are solved in terms of $t_1-t_2$ and $t_1-t_3$. That is to say that we have two relations amongst these four variables, so only two of them are free.  This uses up two of the eight equations, and leaves us with six equations to solve with only five remaining unknowns, $\theta, a^i$ and $t_1$.  This has generically
no solution.  

It is easy to see that if we increase the numbers of particles to four or more the counting gets rapidly worse. Suppose we add a fourth worldline, which 
we want to make intersect with the first three by the same Lorentz transformation.  We are interested in the generic case, so we assume that in the 
original frame of reference this fourth worldline does not coincide with any of the first three. We want to solve the problem of whether there is a Lorentz
transformation that can make all four worldines coincide at some common event; clearly this only has a solution if that Lorentz transform already 
brings the first three worldlines to coincidence.  So we can count the additional degrees of freedom and additional equations to be solved.  
There is one more degree of freedom, which is the time on the new worldline, $t_4$.  But there are four equations to be solved for each value or
$t_4$ if the event $x^a_4 (t_4)$  is to be brought by the Lorentz transformation to the intersection point the other three are brought to.  So there are
a total of three more constraints for a total of ten equations to be solved by six free variables. The situation is that $\kappa$-intersecting quadruples of worldlines are even less generic.

\subsection{The case of $d=1$ spatial dimension}

The  analysis in \cite{sabine-paradox} is carried out in the $1+1$ dimensional case.  We have to analyze that case separately because there is a special situation, which is that
every two non-parallel worldlines meet somewhere.  This means every triplet of worldlines has three generically three intersection points where
they meet in pairs.   Let us call the events where worldlines $x_{\alpha}^a (\tau)$ and $x_{\alpha}^a (\tau)$ meet $z_{\alpha \beta}^a$.

The phase space is initially six dimensional.

Let us start with the case where three wordlines intersect at a single point, 
\f
p^a = z_{12}^a= z_{23}^a =z_{13}^a.
\label{3meetin2}
\ff
The condition that all
three intersection points are equal is four equations.  So the set of histories in which this is the case is a $2$ dimensional subspace of the phase space.
Indeed, it is parameterized by the points $p^a$ of the intersection, which is two dimensional. 

Now the set of Lorentz boosts is in this case $3$ dimensional, one for the single boost parameter, and two for the origin around which the boost is made.  
However of the two degrees of freedom that come from translating the boosts, we have already seen from (\ref{ttrans}) that only one comes into the
energy dependence of the Lorentz transformations.  So we have four equations to solve with two parameters.   (The freedom to choose the $\tau_\alpha$ is taken into account here by the fact that there are generic intersection points of each pair we are trying to bring together at one event.)

So the question is, can the four equations that represent the coincidence of the three intersection points be generated from a generic point by the two degrees of freedom of Lorentz boosts.  The answer is generically no.  That is, consider a generic history with three worldlines that meet in pairs.
This is a six dimensional set.  They cannot be all brought to an element of the two dimensional set where all three coincide at a single point by a three parameter family of Lorentz boosts.  

So we see that even in $d=1$ dimensions having three  $\kappa$-intersecting worldlines are a non-generic set of measure zero.  

There is another way to see this.  If we look at (\ref{xtrans}) for $d=1$ we see that the term in $L^i$ drops out and there is just the ordinary 
Lorentz transformation, which is not energy dependent.  The only energy dependence in the Lorentz transformations is the term
(\ref{ttrans2}) we have already talked about.  So consider three generic worldlines, which intersect in pairs to give three intersection points.  The only thing that the energy dependence of the Lorentz transformations can do is to move the three worldlines up or down by (\ref{ttrans2}).  The three worldlines
move by different amounts due to the different $E_\alpha$'s, but all three are determined by one parameter which is $\omega \cdot a$.  The condition
that the three intersection points shrink to one is four equations, but one only has a single parameter to tune to satisfy them.  Generically there will
be no choice of  $\omega \cdot a$ that reduces the three intersection points to a single triple intersection point.  

\subsection{Stars and so forth}

One might worry that the non-localities inherent in $DSR$ theories will change the bulk properties of large collections of matter such as stars in a way that causes our observations of them to disagree with theory\footnote{Sabine Hossenfelder, personal communication.}.  
That is, even if $\kappa$-intersecting triples of worldlines are rare, there are
so many triples of atoms in stars that some of them will be sometimes $\kappa$-local, so some $\kappa$-local interactions will take place.

Let us grant this for the moment. The problem is, how this is to be observed?  A star is already in thermal equilibrium from a myriad of local interactions.
These have mainly the effect of reproducing the thermal distribution of the atoms in the star, which means they are completely randomized.  If we throw in a small proportion of non-local interactions this is not going to have a measurable effect on the state of thermal equilibrium. Since the $\kappa$-local interactions are, by the postulate of observer independence, of the same kinds as local interactions, nothing is going to happen that doesn't already happen much more numerously with local interactions.  They are not going to change the temperature, pressure or equation of state of the interior of the star by a measurable amount. 

There are in fact models of statistical mechanical systems on lattices that are modified by the addition of a small number of 
non-local interactions\cite{smallworld}.  
There are observable effects when the proportion of non-local effects are high enough, but this is very unlikely to be the case here.  
Moreover the effects that do occur are subtle and not easily measured, one for example is a slight change in the temperature of phase transitions.  
It is hard to imagine making a measurement that could detect non-local interactions through a small change in the temperature of some phase transition in the nuclear matter making up the star, as this is already likely to be in phenomena already parameterized by some phenomenological model of the equation of
state of the gas or nuclear matter making up the star or neutron star. 

In any case, to study the possibility of observable effects due to $\kappa$-locality is a research problem, which would be interesting to carry out.  Given that the result is very unlikely to be significant, because that 
$\kappa$ intersecting triples are a set of measure zero, it would be premature to speculate that any observable effect is produced.

\section{Conclusions}

The main aim of this paper has been to show how paradoxes of locality can be generated in at least one approach to $DSR$-that based on
free particles in $\kappa$-Minkowski spacetime-but only at the classical level, and to propose  that when the quantum dynamics are taken into account those paradoxes may be resolved.   I was able to provide partial evidence, but not proof, in support of this hypothesis. 

I also investigated the possibility that the notion of locality be generalized to a notion appropriate to the physics of classical particles described by
worldlines in the non-commutative geometry of $\kappa$-Minkowski spacetime. 

The conclusions can, in more detail, be stated as follows.   

\begin{itemize}

\item{}In section 3.4 we showed how to generate classical paradoxes of locality, by using the dependence of the Lorentz transformations on energy, momentum and angular momentum.   We considered a case of two or more particles which coincide in one observers frame, and found expressions for their
differences in space and time coordinates as seen in a second frame, given by equations (\ref{differencex}) and (\ref{differencet}).  

\item{}  In section 4.4.1 we set up the quantum theory for a free relativistic particle in $\kappa-$Minkowski spacetime, computed the propagation of a Gaussian wavepacket, defined initially in momentum space, transformed to
spacetime, and found that there is an anomalous spreading of the wavepacket given by (\ref{Deltaw}).  This is proportional to 
$l_p x \Delta p/\hbar$.

Thus, we can conclude that the classical paradox is resolved for quantum states in which $\Delta p \geq |p|$.  On the other hand, if
$\Delta p \leq |p|$ then the paradoxes may remain.  What is the meaning of this?

Since we are working with a gaussian wavepacket, and we can assume that $E << E_p$,  we find that the paradoxes are resolved
so long as $\Delta x < \lambda$, where $\lambda = \hbar / p$ is the wavelength of light.  This means that the particle picture is preferred
to the wave picture because the photon (or massless particle) can be localized to within its wavelength.  We may argue that this is the regime where a single propagating photon can be treated as a quantum free relativistic particle, which is the approximation we are employing here.  


We can  speculate that the other case, where $\Delta x >  \lambda$ requires the full treatment of the quantum field theory to resolve.  
We remark that there are known to be additional sources of non-locality coming from interactions in non-commutative field theory, 
which may play a role in the resolution of
the paradoxes in the full interacting quantum field theory. We may hypothesize that a notion like $\kappa$-locality may be relevant there.

\item{} I proposed in section 5 a weakening of locality to $\kappa$-locality defined
by worldlines intersecting in the frame of some inertial observer.  I showed that this property is very non-generic, so it is  the case that only a 
set of measure zero of $N \geq 3$ worldlines  $\kappa$-interact.  

\end{itemize}

So, to conclude, we have shown that there is a regime where quantum effects suffice to resolve
the paradoxes, but we have not so far shown this in general.   

We note that the problem of measurements and uncertainty in $\kappa$-Minkowski spacetime has been studied before\cite{ALN},
and the effect of spreading of wavepackets has been noted\cite{Rychkov,Tezuka}.  However these previous papers do not seem to have studied propagation
in the $(x^i, T)$ commuting coordinates.  The novel assertion of the present work then appears to be the insistence that the spacetime
we experience must be a commutative spacetime, which means that quantum state evolve in space according to the time coordinate
$T$ which is different from the time coordinate $t$ in which we can study evolution in momentum space.

There are in addition a number of questions and issues that remain to be sorted out in this area.

\begin{itemize}

\item{} The commutation relations (\ref{tx0}) and (\ref{Tp0}) are not invariant under time reversal ($t \rightarrow -t$ or $T \rightarrow -T$, respectively.
Another way to say this is that the theory seems physically different depending on whether the sign of $t_p$ is taken negative, or positive as we have
taken it here.  Are there physical consequences of this?

\item{}Similarly, the theory does not seem invariant under the exchange of $E$ for $-E$.  Is there still $CPT$ symmetry?

\item{}It appears that the phase velocity computed above is not equal to the classical computation of $dx^i/dT$ computed via
the Hamiltonian dynamics.  Is this a problem?  

\item{}It is known that $QFT$ defined on noncommutative geometries in which time and space coordinates don't commute are not unitary\cite{notunitary}.
Might it be that unitary evolution could be defined instead in the time variable $T$?

\item{}There is a long standing debate as to whether the physics in $\kappa$-Minkowski spacetime should be invariant, not just under Lorentz transformations, but under non-linear redefinitions of the phase space coordinates.   A view that seems reasonable is that $DSR$ theories will be limits of quantum theories
of gravity, whose physics will define the appropriate notion of locality and hence the physical spacetime. Thus, a particular set of choices for physical momenta, energy and spacetime coordinates will be reflected in the form of the Hamiltonian. At the same time, any quantum mechanical system is invariant under changes of basis in Hilbert space.  It may be that the observer independence provided by choices of basis in Hilbert space implies an expansion of the notion of observer independence in a quantum theory of gravity, but this is a subtle issue that goes beyond the scope of this paper.  

\end{itemize}

Finally, it is important to remark that even if the proposed apparent paradoxes are in the end resolved, the result is extremely important because the quantum effects necessary to resolve them introduce a stochastic influence of energy on arrival times.  This is relevant for experiments that test the possible energy dependence of the speed of light.  For this reason the paradoxes proposed in \cite{Unruh-DSR,sabine-paradox} are playing a central role in the development of our understanding of the possible observational consequences of quantum gravity. 

\section*{ACKNOWLEDGEMENTS}

This paper was motivated by the hope to answer the  paradoxes proposed by Sabine Hossenfelder\cite{sabine-paradox} and I am grateful
for many discussions and correspondence with her.  I  also have to thank her for pointing out some errors in a previous version of this paper which
were embarrassing, even if they did not affect the conclusions.  
 Giovanni Amelino-Camelia and Jurek Kowalski-Glikman have been extremely helpful commenting on drafts of the manuscript and correcting my misunderstandings about the $\kappa$ world.  Conversations and correspondence with Laurent Freidel,  Joao Magueijo, Seth Major and Chanda Prescod-Weinstein have also been very helpful.  I am also grateful to Giovanni Amelino-Camelia for forwarding to me a draft of a paper which addresses related issues from a different perspective\cite{GTetal}. 
Research at Perimeter Institute for Theoretical Physics is supported in part by the Government of Canada through NSERC and by the Province of
Ontario through MRI.

\end{document}